\documentclass[a4paper,11pt]{article}
\usepackage{cite}
\usepackage{graphicx}
\def\Vec#1{\mbox{\boldmath $#1$}}
\begin{document}
\Large
\centerline{\Large \bf Magnetization Plateau in the $\mbox{\boldmath $S=1$}$ Spin Ladder}
\centerline{\Large \bf with Competing Interactions}

\bigskip

\large

\centerline{\large Nobuhisa {\sc Okazaki}$^{\rm a,*}$, Kiyomi {\sc Okamoto}$^{\rm b}$ and
T\^oru {\sc Sakai}$^{\rm c}$}

\normalsize
\bigskip

\centerline{\large \it $^{\rm a}$ Kyoto Miyama High School, Murashita, Sasari, Miyama-cho,}
\centerline{\large \it  Kitakuwata-gun, Kyoto 601-0705, Japan}
\centerline{\large \it $^{\rm b}$Department of Physics, Tokyo Institute of Technology,}
\centerline{\large \it Oh-okayama, Meguro-ku, Tokyo 152-8551, Japan}
\centerline{\large \it $^{\rm c}$Tokyo Metropolitan Institute of Technology,}
\centerline{\large \it Asahigaoka, Hino, Tokyo 191-0065, Japan}

\bigskip
\bigskip


Very recently a non-trivial magnetization plateau at 1/4 of the
saturation magnetization was observed in the $S=1$ spin ladder
BIP-TENO.
In our previous work we proposed a possible mechanism of the plateau
based on the second- and third-neighbor exchange couplings which lead to
frustration.
In order to confirm the realization of the mechanism,
we compare the temperature dependence of the magnetic susceptibility and
some
critical magnetic fields obtained by the numerical calculation for the
proposed model with the experimental results.

\bigskip
\bigskip
\noindent

\bigskip
\noindent
*Corresponding author: \\
Nobuhisa Okazaki \\
Email: nobuhisa@sci.himeji-tech.ac.jp \\
Fax:+81-771-77-0821

\newpage

\section{Introduction}

A recent synthesized organic $S=1$ spin ladder, 
3,3',5,5'-tetrakis({\it N-tert}-butylaminoxyl) biphenyl, 
abbreviated as BIP-TENO\cite{KK}, is one of interesting strongly correlated 
electron systems. 
It exhibits a field-induced spin gap which is observed as a 
plateau in the magnetization curve. 
The high-field measurement\cite{TG} indicated that the plateau appears 
at 1/4 of the saturation moment. 
Such a magnetization plateau is predicted in various systems
\cite{cabra1,cabra2,totsuka,tonegawa,
tonegawa-okamoto,st}.  
A general condition of the quantization of the magnetization was derived 
from the Lieb-Schultz-Mattis (LSM) \cite{lsm} theorem 
for low-dimensional magnets \cite{oshikawa}.
The necessary condition of the plateau is 
was described as 
\begin{eqnarray}
Q(S-m) = {\rm integer}
\label{quantization} 
\end{eqnarray}
where $Q$ is the spatial period of the
 ground state measured by the unit cell. 
$S$ and $m$ are the total spin and the magnetization per unit cell, 
respectively. 
Applying this theorem to the BIP-TENO, 
the 1/4 plateau is the case of 
$S=2$ and $m=1/2$.
Therefore, a spontaneous breaking of the translational symmetry(maybe $Q=2$) 
must occur at the plateau.
In the previous work \cite{Oos} 
by the present authors two mechanisms of the 1/4 
plateau of the $S=1$ spin ladder were proposed, 
based on the frustrated interactions. 
In the next section we briefly review the mechanisms and show the 
phase diagrams obtained by the level spectroscopy analysis \cite{no}. 
The main purpose of this paper is 
to consider the realization of the mechanism 
at the 1/4 plateau of BIP-TENO, with some quantitative analyses on 
the critical magnetic fields and the temperature dependence of the 
susceptibility.

\section{Mechanisms of 1/4 plateau}

As an origin of the 1/4 magnetization plateau in the $S=1$ spin ladder, 
we introduce the second and third-neighbor exchange interactions. 
The model is 
described by the Heisenberg Hamiltonian
\begin{eqnarray}
{\hat H}&=&{\hat H}_0+{\hat H}_Z \\
{\hat H}_0&=&J_1\sum_i^L({\Vec S}_{1,i} \cdot {\Vec S}_{1,i+1}
+{\Vec S}_{2,i} \cdot {\Vec S}_{2,i+1}) 
+J_{\perp}\sum_i^L {\Vec S}_{1,i} \cdot {\Vec S}_{2,i} \nonumber\\
& &+J_2\sum_i^L({\Vec S}_{1,i}
\cdot {\Vec S}_{2,i+1}+{\Vec S}_{2,i} \cdot {\Vec
S}_{1,i+1}) \nonumber \\
& &+J_3\sum_i^L({\Vec S}_{1,i}
\cdot {\Vec S}_{1,i+2}+{\Vec S}_{2,i} \cdot {\Vec
S}_{2,i+2})\\
{\hat H}_Z&=&-H\sum_i^L({S_{1,i}^z}+{S_{2,i}^z}),
\label{ham}
\end{eqnarray}
where $J_{1},J_{\perp},J_2$ and $J_3$ denote the coupling constants 
of the leg, rung and second- (diagonal) and third- exchange interactions,
respectively (Fig. 1). 

\begin{figure}
  \begin{center}
     \scalebox{0.3}[0.3]{\includegraphics{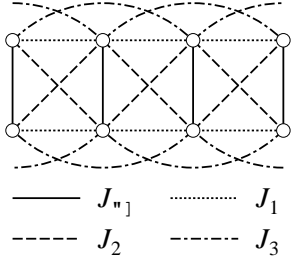}}
     \caption{Spin ladder with second and third exchange interactions.}
  \end{center}
\end{figure}

Hereafter we put $J_{\perp}$=1.
${\hat H}_Z$ is the Zeeman term where $H$ denotes the magnetic field
along the $z$-axis and the eigenvalue $M$ of the conserved quantity
$\sum_{i}{({S_{1,i}^z}+{S_{2,i}^z})}$
is a good quantum number.
The macroscopic magnetization is represented by $m=M/L$.
In this definition the 1/4 of the saturation magnetization 
corresponds to $m=1/2$. 
In order to explain the mechanism of the plateau at $m=1/2$, 
we use the degenerate perturbation theory around 
the strong rung coupling limit $J_1,J_2,J_3 \ll 1 $ \cite{oos,mila}.
For the two spins at each rung, 
we take only two dominant states; the singlet 
$\Psi_{0,0} \equiv (|\uparrow \downarrow \rangle 
-|\downarrow \uparrow \rangle$ 
and the triplet $\Psi_{1,1}\equiv |\uparrow \uparrow \rangle $.
We introduce a pseudo spin $\Vec T$ for each rung coupling and map the two
original sates singlet $\Psi_{0,0}$ and triplet $\Psi_{1,1}$ of the $\Vec S$
picture to the $|\Downarrow \rangle $ and $|\Uparrow \rangle $ states of
$\Vec T$, respectively. 
Effective Hamiltonian in pseudo spin can be written as follows:

\begin{eqnarray}
{\hat H}_{\rm eff} =&&
\frac{8(J_1-J_2)}{3}\sum_i^L({T}^x_{i} \cdot {T}^x_{i+1}
+{T}^y_{i} \cdot {T}^y_{i+1}) \nonumber\\
&&+\frac{J_1+J_2}{2} \sum_i^L({T}^z_{i} \cdot {T}^z_{i+1}) 
+\frac{8J_3}{3}\sum_i^L({T}^x_{i} \cdot {T}^x_{i+2}
+{T}^y_{i} \cdot {T}^y_{i+2})  \nonumber\\
&&+\frac{J_3}{2}\sum_i^L({T}^z_{i} \cdot {T}^z_{i+2}).
\label{eham}
\end{eqnarray}
This is the Hamiltonian of the $T=1/2$ $XXZ$ chain with the second-neighbor 
interaction. 
The magnetization $m=1/2$ of the original system corresponds to 
$m=0$ in the pseudo-spin system. 
Referring the well-known features of the $S=1/2$ frustrated $XXZ$ chain \cite{no}, 
sufficiently large $J_2$ and $J_3$ lead to the N\'eel order and 
dimerization of the pseudo spins, respectively. 
They correspond to the field-induced spin gap at $m=1/2$, 
that is the 1/4 plateau in the original system. 
The boundary between the spin-fluid and plateau phases is of the 
Kosterlitz-Thouless (KT) type \cite{kt}.
Therefore, the pseudo-spin picture gives 
two different mechanisms of the plateau; N\'eel order and dimerization, 
denoted as plateaux $A$ and $B$, respectively. 

The KT phase boundary can be determined precisely, 
using the recently developed level spectroscopy \cite{no} applied 
to the low-lying energy levels of the finite chains obtained by 
the numerical diagonalization. 
We show only thus-obtained phase diagrams in the $J_2$-$J_1$
($J_3=0$) and $J_3$-$J_1$($J_2=0$) planes in Figs. 2 and 3, 
respectively. 
The two gapless phases in Fig. 2 correspond to two different 
ordered states in the classical limit,  
depending on whether the rung or diagonal
interactions are dominant \cite{so}.
Note that there is an upper bound of $J_1$($\sim 0.7$) for the 
plateau $A$, while no bound for the plateau $B$. 

\begin{figure}[h]
      \begin{center}
         \scalebox{0.4}[0.4]{\includegraphics{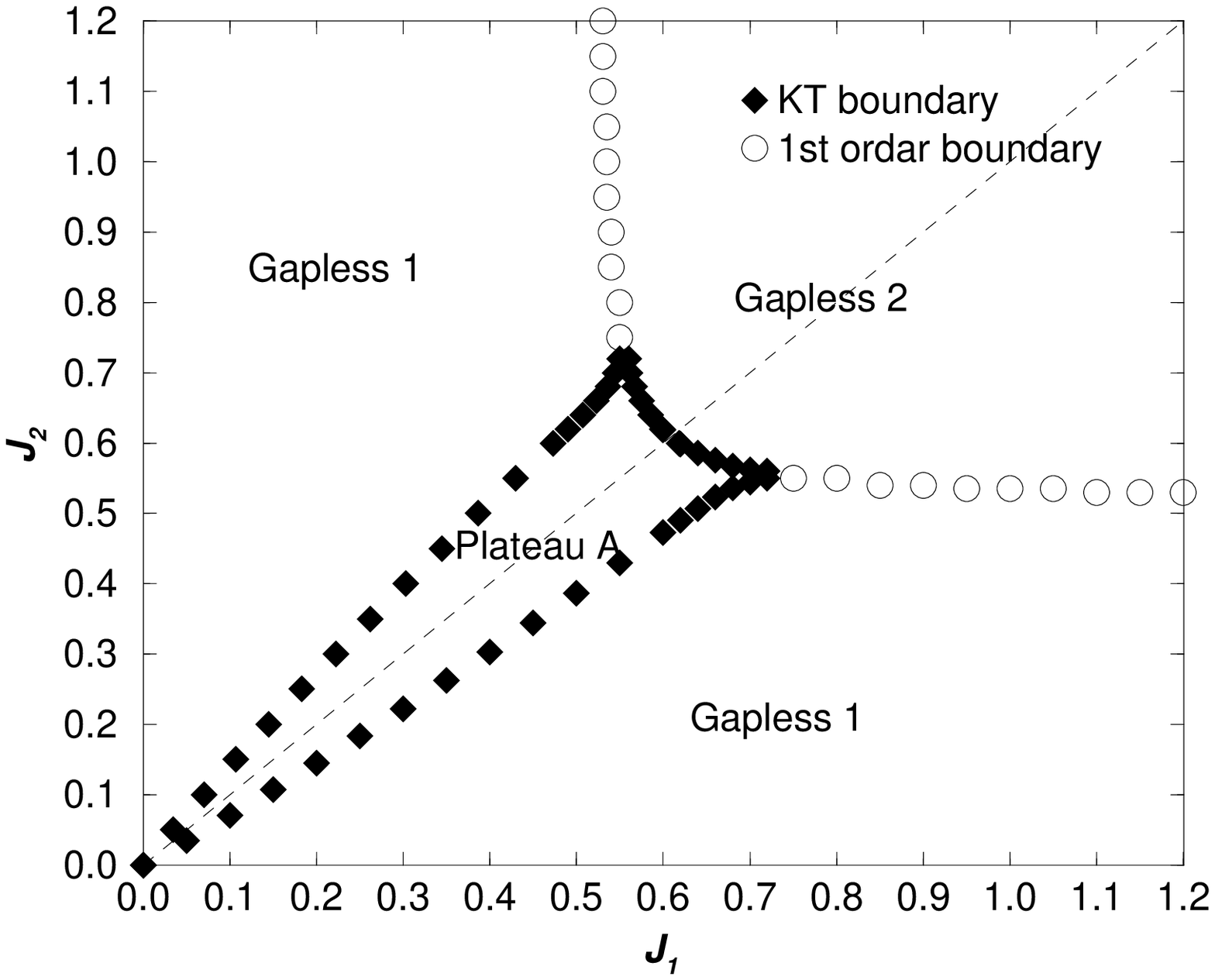}}
      \end{center}
      \caption{Phase diagram on the $J_2$-$J_1$ plane at $m=1/2$}
      \label{fig:level-cross}
      \begin{center}
         \scalebox{0.4}[0.4]{\includegraphics{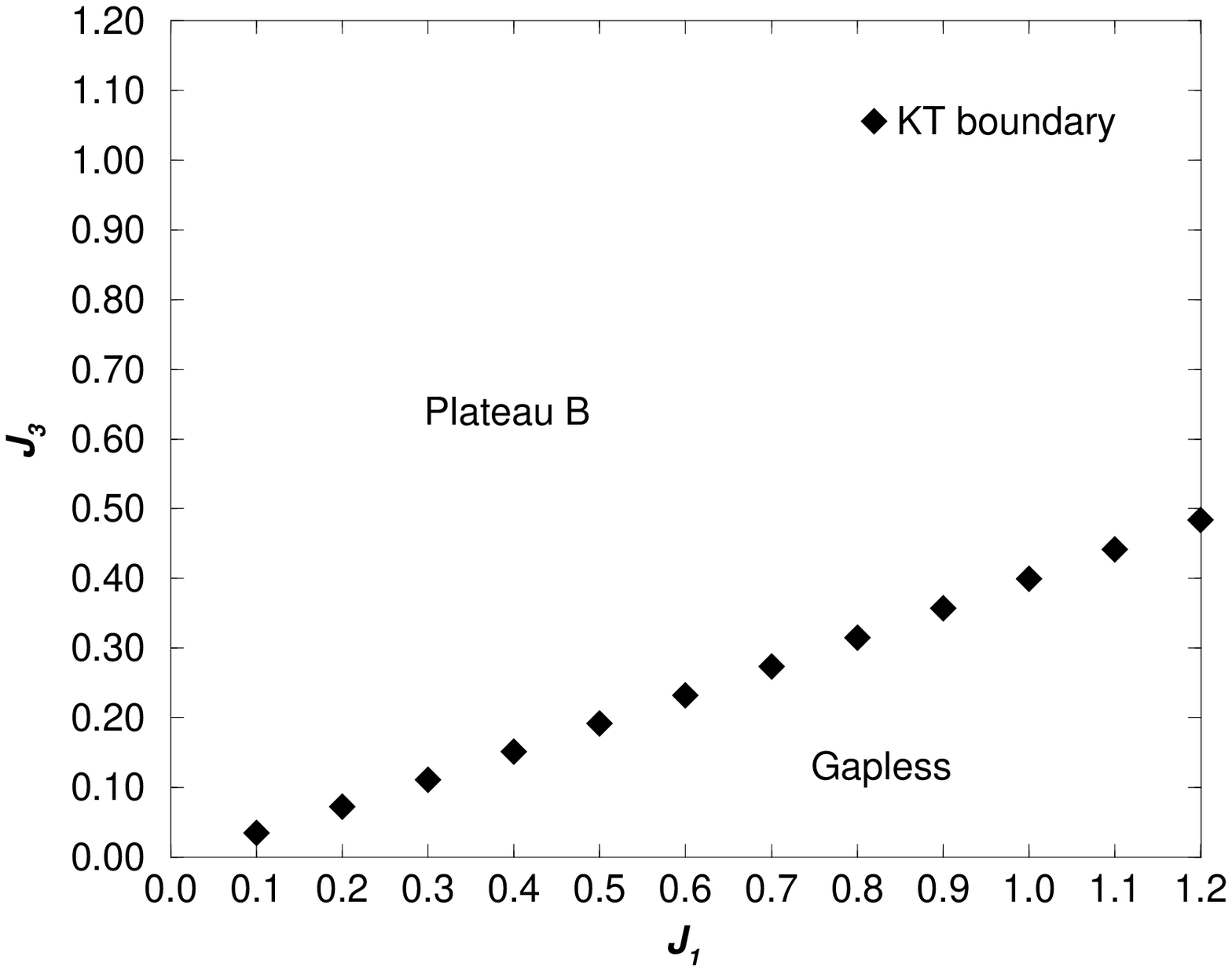}}
      \end{center}
      \caption{Phase diagram on the $J_3$-$J_1$ plane at $m=1/2$.}
      \label{fig:phase-diagram}
\end{figure}

\section{Comparison with Experiment}

Based on the obtained phase diagrams, 
we discuss the realistic mechanism of the 1/4 plateau of BIP-TENO. 
The ratio $J_1/J_{\perp}$ of BIP-TENO was estimated as 
$J_1/J_{\perp}\sim 1.2$, fitting the observed temperature dependence 
of the susceptibility $\chi $ to the numerical calculation for the 
$S=1$ simple spin ladder \cite{KK}. 
Fig. 2 suggests that there is no chance of the N\'eel plateau 
for $J_1/J_{\perp}\sim 1.2$. 
In addition the required value of the $J_2/J_1$ for the plateau 
is about 0.69 even for $J_1/J_{\perp} < 0.7$. 
$J_1/J_{\perp} \sim$ 0.69 is too large for the realization. 
Thus the N\'eel mechanism due to $J_2$ should be discarded. 

Next we consider the possibility of the dimer plateau due to $J_3$. 
According to Fig. 3, 
the plateau would appear if $J_3/J_1 >0.39$ for $J_1/J_{\perp} \sim 1.2$. 
$J_3/J_1 \sim 0.39$ is not so far from the realization, 
because the lattice spacing along the leg is much smaller than the rung 
in the crystal structure of BIP-TENO \cite{KK}. 
Thus we examine the dimer mechanism due to $J_3$ more quantitatively. 

\begin{figure}[h]
      \begin{center}
         \scalebox{0.4}[0.4]{\includegraphics{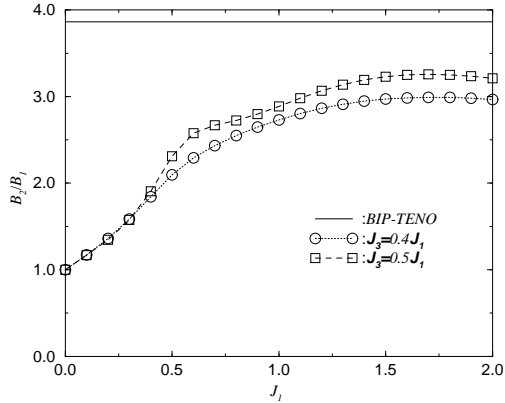}}
      \end{center}
      \caption{The ratio of the critical magnetic field ($B_2/B_1$).}
      \label{fig:level-cross}
\end{figure}

Using the numerical diagonalization of finite clusters up to $L=8$ (16 spins) and 
the size scaling technique on the model (2), 
we estimate the ratio of the critical magnetic field $B_1$ that
spin gap disappears and $B_2$ that plateau begin to appear. 
The results for $J_3/J_1$=0.4 and 0.5 are plotted versus $J_1$ 
in Fig. 4, gathered with the experimental result of BIP-TENO.  
It suggests that the calculated $B_2/B_1$ is closer to that of BIP-TENO 
for $J_1/J_{\perp}\sim 1.7$, rather than for $J_1/J_{\perp} \sim 1.2$. 
However the experimental estimation $J_1/J_{\perp} \sim 1.2 $ is not 
so conclusive, because the fitted curve was not obtained by the 
numerical diagonalization of the $S=1$ ladder, but by some 
mean field approximation for the rung interaction. 
Thus it would be important to fit the direct numerical calculation 
for the $S=1$ ladder including $J_3$ to observed $\chi$ for 
$J_1/J_{\perp} =1.7$. 
We performed the finite-temperature Lanczos method \cite{jaklic} 
to calculate the temperature dependence of $\chi$ for the 
system $L=8$. 
The results for $J_1/J_{\perp}=1.7$ and various values of $J_3/J_1$ 
are shown in Fig. 5, together with the experimental result of BIP-TENO. 
The numerical curve well agrees with the measured one for $J_3/J_1$=0.4 
or 0.5. 
Therefore, the dimer plateau is expected to realize, 
if $J_1/J_{\perp}\sim 1.7$ and $J_3/J_1 \sim $0.4 or 0.5 are satisfied 
in BIP-TENO. 
A little difference in $B_2/B_1$ between the model calculation and 
the experiment is possibly due to the finite size effect, 
because we used only the values for $L=4$ and 8 to estimate it. 
Indeed the ratio $B_2/B_1$ of the finite systems is revealed to 
increase with increasing $L$ toward the experimental result. 

\begin{figure}[h]
      \begin{center}
         \scalebox{0.4}[0.4]{\includegraphics{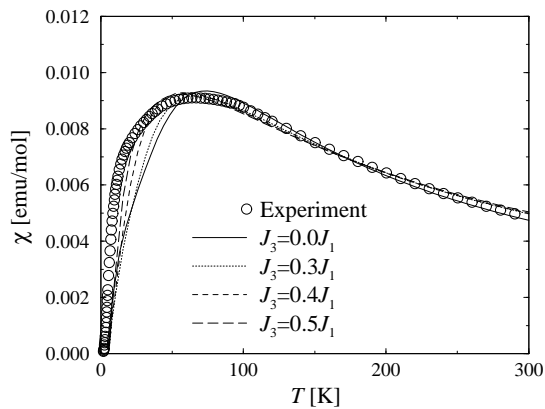}}
      \end{center}
      \caption{The temperature dependence of the susceptibility $\chi$.}
      \label{fig:phase-diagram}
\end{figure}

\section{summary}

We proposed two mechanisms of the 1/4 magnetization plateau in 
the $S=1$ frustrated spin ladder. 
They are described by the N\'eel order and dimerization of the 
pseudo-spin system. 
Comparing the ratio of the two critical fields $B_2/B_1$ and 
the temperature dependence of $\chi$ between the model calculation 
and the experiment, 
we conclude that the dimer mechanism due to the third neighbor 
interaction is more suitable for BIP-TENO.

\section*{Acknowledgements}
We wish to thank Professor Tsuneaki Goto for sending some experimental 
data of BIP-TENO.

\end{document}